\documentclass[floats,floatfix,amssymb,prl,twocolumn,superscriptaddress,nofootinbib]{revtex4-1}
		
\usepackage{amssymb,amsmath,verbatim,mathtools,needspace,enumitem,etoolbox,graphicx,physics,microtype,afterpage,bm}
\usepackage[dvipsnames, usenames]{xcolor}
\definecolor{linkcolor}{rgb}{0.0,0.3,0.5}
\usepackage[unicode, colorlinks=true, linkcolor=linkcolor, citecolor=linkcolor, filecolor=linkcolor,urlcolor=linkcolor, pdfusetitle]{hyperref}
\usepackage[all]{hypcap}
\usepackage[T1]{fontenc}
\usepackage[utf8]{inputenc}
\usepackage{tabularx}
\usepackage{float}
\renewcommand{\arraystretch}{1.4}

\usepackage{lmodern}
\allowdisplaybreaks
\usepackage{tikz}
\usepackage{color}
\usepackage{framed}
\usepackage{hyperref}
\hypersetup{colorlinks, citecolor=bluscuro, linkcolor=black, urlcolor=bluscuro}
\definecolor{rossos}{cmyk}{0,1,1,0.55}
\definecolor{bluscuro}{rgb}{0.15, 0.2, .85}
\definecolor{bluchiaro}{cmyk}{1,.3,0.,0.1}
\definecolor{ForestGreen}{rgb}{0.13, 0.55, 0.13}

\newcommand{\be}{\begin{equation}}
\newcommand{\ee}{\end{equation}}
\newcommand{\bea}{\begin{equation}\begin{aligned}} 
\newcommand{\eea}{\end{aligned}\end{equation}}
\renewcommand{\d}{{\rm d}}

\newcommand{\lp}{\left (}
\newcommand{\rp}{\right )}

\def\lsim{\mathrel{\rlap{\lower4pt\hbox{\hskip0.5pt$\sim$}}
    \raise1pt\hbox{$<$}}}         
\def\gsim{\mathrel{\rlap{\lower4pt\hbox{\hskip0.5pt$\sim$}}
    \raise1pt\hbox{$>$}}}         

\def\d{{\rm d}}

\def\PBH{\text{\tiny PBH}}
\def\rad{\text{\tiny rad}}
\def\eq{\text{\tiny eq}}

\def\cl{\text{\tiny cl}}

\def\E{\text{\tiny  E}}
\newcommand{\ie}{{\it i.e.}}
\newcommand{\eg}{{\it e.g.}}

\begin{document}

\title{Ruling out Initially Clustered Primordial Black Holes as Dark Matter}

\author{Valerio De Luca}
\email{valerio.deluca@unige.ch}
\affiliation{D\'epartement de Physique Th\'eorique and Centre for Astroparticle Physics (CAP), Universit\'e de Gen\`eve, 24 quai E. Ansermet, CH-1211 Geneva, Switzerland}

\author{Gabriele Franciolini}
\email{gabriele.franciolini@uniroma1.it}
\affiliation{Dipartimento di Fisica, Sapienza Università 
	di Roma, Piazzale Aldo Moro 5, 00185, Roma, Italy}
\affiliation{INFN, Sezione di Roma, Piazzale Aldo Moro 2, 00185, Roma, Italy}

\author{Antonio Riotto}
\email{antonio.riotto@unige.ch}
\affiliation{D\'epartement de Physique Th\'eorique and Centre for Astroparticle Physics (CAP), Universit\'e de Gen\`eve, 24 quai E. Ansermet, CH-1211 Geneva, Switzerland}

\author{Hardi Veermäe}
\email{hardi.veermae@cern.ch}
\affiliation{NICPB, Rävala pst. 10, 10143 Tallinn, Estonia}


\begin{abstract}
\noindent
Combining constraints from microlensing and Lyman-$\alpha$ forest, we provide a simple argument to show that large spatial clustering of stellar-mass primordial black holes at the time of formation, such as the one induced by the presence of large non-Gaussianities,  is ruled out.
Therefore, it is not possible to evade existing constraints preventing stellar-mass primordial black holes
to be a dominant constituent of the dark matter by boosting their initial clustering.
\end{abstract}

\maketitle

\noindent{{\bf{\it Introduction.}}}
The physics of Primordial Black Holes (PBHs) has attracted a lot of interest~\cite{Sasaki:2018dmp, Carr:2020gox, Green:2020jor} thanks to the multiple detections of gravitational waves (GWs) 
coming from BH binary mergers~\cite{LIGOScientific:2016aoc, LIGOScientific:2018mvr,LIGOScientific:2020ibl, LIGOScientific:2021djp} and the suggestion that some of them may be of primordial origin \cite{Bird:2016dcv,Sasaki:2016jop,Clesse:2016vqa}. 

One of the fundamental questions about PBHs is whether they can contribute significantly to the dark matter (DM) abundance. In the LIGO-Virgo-KAGRA (LVK) mass range, the answer seems to be negative. Both microlensing data~\cite{Carr:2020gox}, as well as an otherwise too high merger rate~\cite{Ali-Haimoud:2017rtz,Hutsi:2020sol, DeLuca:2021wjr, Franciolini:2021tla}, impose the fraction $f_\PBH$ of PBHs in DM to be below the percent level. These constraints are derived assuming that PBHs are initially Poisson distributed in space. PBH clustering at the time of formation can, in principle, change the present and past PBH merger rate by both boosting the formation of binaries and increasing the subsequent suppression due to the interaction of binaries in PBH clusters, thus possibly allowing for larger values of $f_\PBH$~\cite{Raidal:2017mfl, Vaskonen:2019jpv, Young:2019gfc, Atal:2020igj, Jedamzik:2020ypm, DeLuca:2020jug}. Similarly, a sizeable initial PBH clustering might relax the microlensing bounds since compact PBH clusters would act as a single lens that is too massive to be probed by microlensing surveys~\cite{Clesse:2017bsw,Calcino:2018mwh,Carr:2019kxo}.

The aim of this paper is to demonstrate that existing observations do not allow for the totality of DM to consist of stellar-mass PBHs for any amount of initial spatial clustering. The argument is quite simple. To significantly affect constraints stemming from the merger rate of PBH binaries, PBHs must be spatially correlated at the kpc comoving scales which are relevant for the present merger rate. Such PBHs will form compact clusters during radiation domination, which are initially Poisson distributed at larger scales, of the order of Mpc. Thus,  existing Lyman-$\alpha$ constraints provide an upper bound on the size of such a PBH correlation. This upper bound is incompatible with the lower bound on the same physical quantity necessary to avoid the microlensing constraints. 

We proceed now to summarise our arguments. Technical details are contained in the {\it Supplemental Material} (SM).

\vskip 0.5cm
\noindent
\noindent{{\bf{\it Modelling PBH clustering.}}} 
PBHs are discrete objects and the two-point correlation function of their density contrast $\delta_\PBH$ takes the general form \cite{Desjacques:2018wuu}
\be\label{eq:delta_2P}
\langle \delta_\PBH(\vec{r})\delta_\PBH(0)\rangle=\frac{1}{\overline{n}_\PBH}\delta_\text{\tiny D}(r)+\xi_\PBH(r),
\ee
where $r=|\vec{r}|$ is the distance between two PBHs and
\be
    \overline{n}_\PBH \simeq 30 f_\PBH \lp \frac{M_\PBH}{M_\odot} \rp^{-1} {\rm kpc}^{-3}
\ee
is the average PBH number density per comoving volume. 
We will assume a monochromatic PBH population with mass $M_\PBH$.
The first term on the right-hand side represents the Poisson term arising from the discrete nature of PBHs. It is present for any distribution of point-like objects regardless of their clustering. The second term $\xi_\PBH(r)$ is the so-called reduced PBH correlation function. 

We will focus on the standard scenario where PBHs form from the collapse of large overdensities when the corresponding wavelengths re-enter the horizon \cite{Sasaki:2018dmp}. In the absence of primordial non-Gaussianity, PBHs are not correlated and the two-point PBH function is dominated by the Poisson term in the range of initial distances relevant for the calculation of the present merger rate \cite{Desjacques:2018wuu,Ali-Haimoud:2018dau}. This is because the formation of PBHs is a very rare event.
For instance, for a PBH with mass $M_\PBH \sim 10 M_\odot$ it happens only in one over $10^8$
Hubble volumes \cite{MoradinezhadDizgah:2019wjf}. However, if some non-Gaussianity is present and correlations over distances larger than the horizon arise during formation, sizeable spatial correlations between PBHs are possible.

As typically done in the literature, we model initial PBH clustering assuming that the reduced two-point correlator is approximately constant in space and much larger than unity up to some comoving clustering scale $r_\cl$,
\be\label{corr_template}
    \xi_\PBH(r)\simeq
\begin{cases}
    \xi_0 \gg 1 \qquad {\rm for} \qquad r\lsim r_\cl,  \\
    0 \qquad \qquad \ \, {\rm otherwise},
\end{cases}
\ee
where $r_\cl \gtrsim 1\, {\rm kpc}$ and $\xi_0 \gtrsim 1$ in order for clustering to be relevant
for comoving scales of PBH binaries with typical mass about 30$M_\odot$ that merge today (see e.g. Refs.~\cite{Ballesteros:2018swv,Atal:2020igj}).
At smaller scales, the two-point correlator can be even larger, but with the spatial exclusion condition that $\xi_\PBH(r)\simeq -1$ below approximately the comoving Hubble radius at formation time, as distinct PBHs cannot form arbitrarily close to each other. 
As we will see, clustered PBHs would induce Poisson perturbations at much larger scales, corresponding to the average cluster distance, well within the Lyman-$\alpha$ range. The relevance of the precise shape of $\xi_\PBH$ is reduced by the fact that the cluster properties after its gravitational collapse are mostly determined by the average overdensity.

Quite generally, the evolution of PBH clustering follows different stages: 

1) initially, the cosmological horizon is comparable to the size of PBHs, thus each PBH forms independently (however, this can depend on the formation mechanism). Nevertheless, their formation probability can be heavily modulated, causing them to preferably form in superhorizon patches of comoving size $r_\cl$. This sets the initial spatial distribution for PBHs.

2) While the initial PBHs density fluctuates due to the initial clustering as well as the discreteness of PBHs, the resulting fluctuations in the total energy density are tiny right after their formation deep in the radiation dominated era. Thus, the PBHs remain coupled to the Hubble expansion. At this stage $\overline{\rho}_\PBH \ll \delta \rho_\PBH \ll \rho_{\rad}$ and $\delta \rho_\PBH/\rho_\rad \propto a$ (the scale factor) due to the faster dilution of radiation. 

3) When $\delta \rho_\PBH \approx \rho_\rad$, 
PBHs begin to decouple from the expansion, 
causing the gravitational collapse 
and a subsequent violent relaxation of these clusters. Due to the high density contrast, this stage takes place deep in the radiation dominated era \cite{Kolb:1994fi}. The resulting gravitationally bound clusters have an average mass 
$M_\cl \simeq M_\PBH N_\cl$
where $N_\cl$ is the PBH number in the cluster (see the SM for more details)
\be\label{eq:N_cl}
    N_\cl 
    \simeq \overline{n}_\PBH \int \d^3 x  \, \xi_\PBH
    \approx \frac{4\pi}{3}\overline{n}_\PBH \xi_0 r^3_\text{\tiny cl}.
\ee
In clustered scenarios, $N_\cl \gg 1$ should be assumed. For definiteness, we impose
\be
\label{Ncl3}
N_\cl \gtrsim 3: \quad \xi_0 \gtrsim 2.3 \cdot 10^{-2} f_\PBH^{-1}  \lp \frac{M_\PBH}{M_\odot} \rp
\lp \frac{r_\text{\tiny cl}}{\rm kpc} \rp^{-3}. \,
\ee

\begin{figure}[t!]
	\centering
	\includegraphics[width=0.45\textwidth]{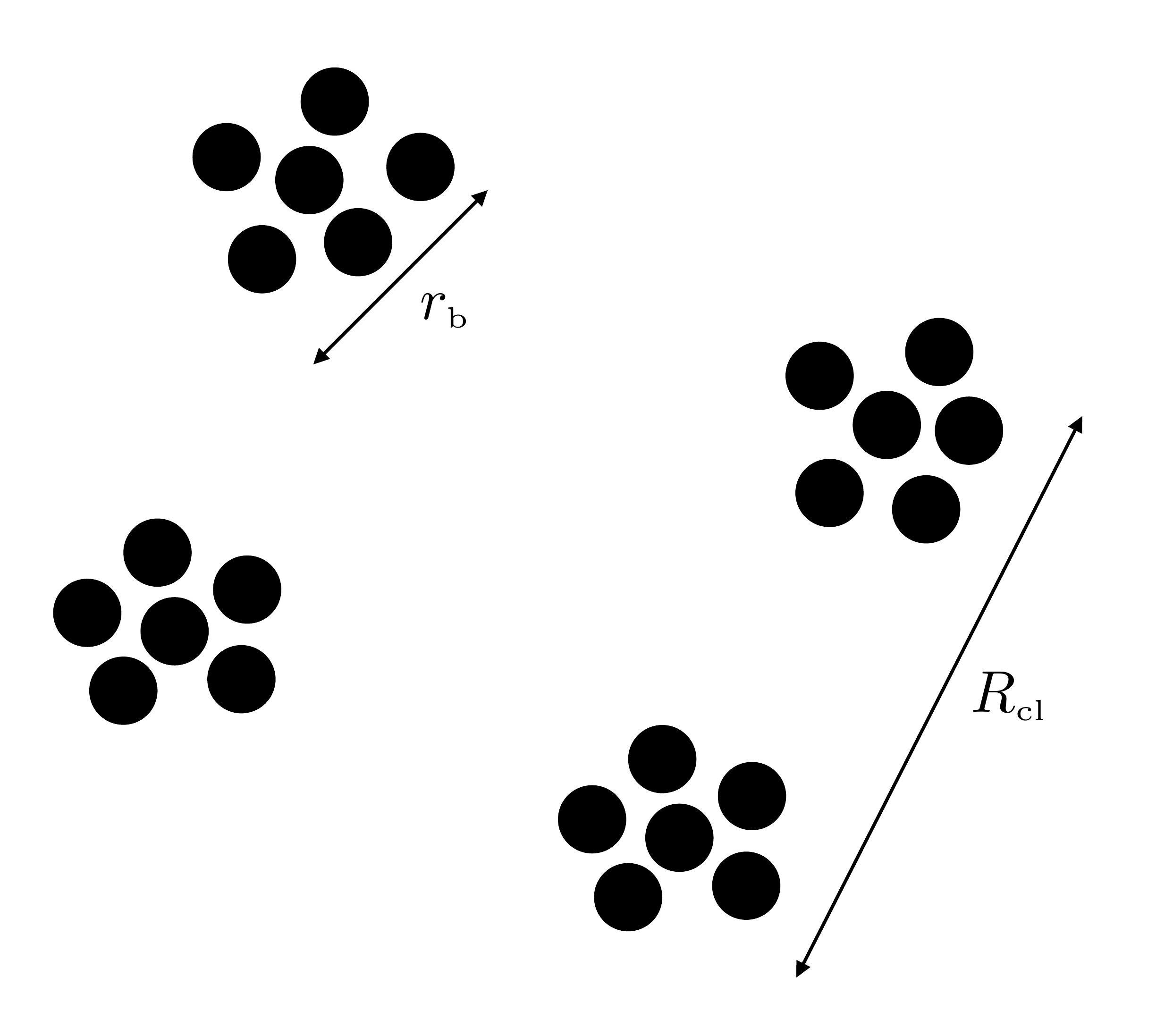}
	\caption{Pictorial representation of the initially clustered PBHs, along with the relevant scales.}
	\label{fig}
\end{figure}  

4) PBH clusters act as point-like objects at comoving distances $R_\cl \gg r_\cl$, where
\be
\label{Rcl}
R_\text{\tiny cl} \simeq \frac{1}{\overline n_\text{\tiny cl}^{1/3}} \simeq \lp \frac{\overline n_\PBH}{N_\text{\tiny cl}} \rp^{-1/3} \simeq r_\text{\tiny cl} \xi_0^{1/3},
\ee
in terms of the cluster number density $\overline n_\text{\tiny cl}$. See Fig.~\ref{fig} for a pictorial representation. Also, clusters themselves begin to group into bound systems after matter-radiation equality. As they are discrete objects, such PBH clusters follow a Poisson distribution and their subsequent evolution will be similar to the early small scale structure formation of PBHs of mass $M_\cl$~\cite{Inman:2019wvr,DeLuca:2020jug}. Note that  Eq.~\eqref{Rcl} requires  $\xi_0 \gg 1$ for the initial clusters not to overlap with each other,
guaranteeing that their gravitational collapse during radiation domination is clearly separated from the 
following Poisson clustering evolution.

The physical radius of the cluster is (see the SM)
\be
   r_\text{\tiny b} \simeq 4 \cdot 10^{-2} {\rm pc} \,
    f_\PBH^{1/3}  
    \xi_0^{-1}
    \lp \frac{C}{200} \rp^{-1/3}
    \lp \frac{r_\text{\tiny cl}}{\rm kpc} \rp,
\ee
in terms of their overdensity $C$ right after decoupling. 
We require that the cluster does not collapse into a heavy PBH of mass $\approx M_\text{\tiny cl}$, as the latter would correspond to a different (but more typical) scenario in which the DM is composed of Poisson distributed heavy PBHs.
Therefore, we demand that the final halo is less compact than a BH, \ie, it must violate the hoop condition $r_\text{\tiny b} \lesssim 2 G M_{\cl}$~\cite{Misner:1973prb}. This is equivalent to demanding that the cluster is much smaller than the cosmological horizon during gravitational collapse or that collapse takes place after the cluster enters the horizon, and translates into
\be
\label{xiBH}
\text{Heavy-PBH}: \quad 
\xi_0 \lesssim 
6 \cdot 10^4\, f_\PBH^{-1/3} \lp \frac{C}{200}\rp^{-1/6} \lp \frac{r_\cl}{\rm kpc} \rp^{-1}.
\ee

\begin{figure}[t]
	\centering
\includegraphics[width=0.48\textwidth]{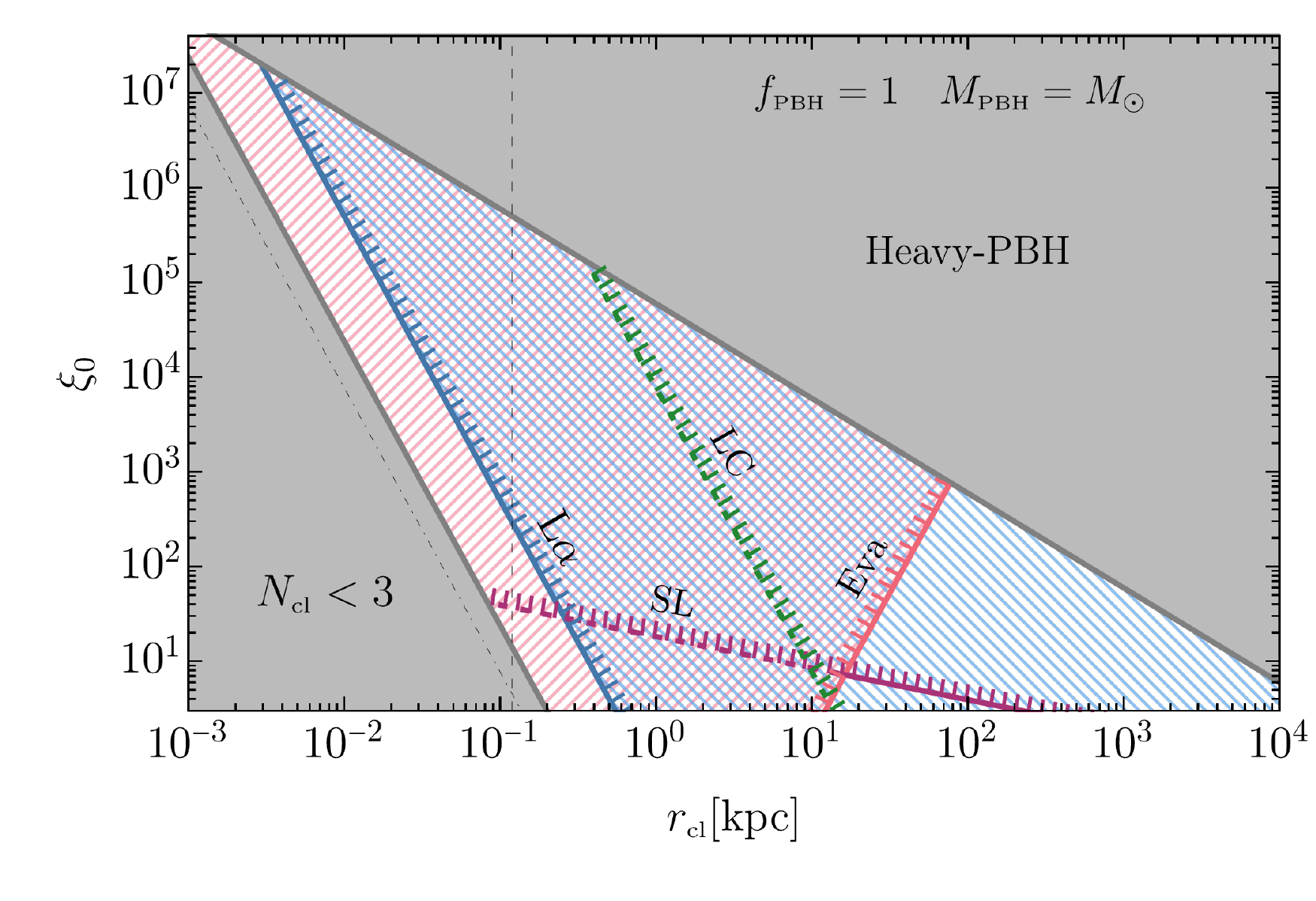}
	\caption{
	Constraints on the relevant parameter space for initial PBH clustering, assuming solar-mass PBHs and $f_\PBH = 1$. 
In gray we shade out regions where the conditions requiring sufficiently large clusters ($N_\cl \gtrsim 3$, Eq.~\eqref{Ncl3}) or avoiding heavy PBH formation (Heavy-PBH, Eq.~\eqref{xiBH}) are not met.
The different lines correspond to 
cluster evaporation (Eva, Eq.~\eqref{xievap}), 
single lens (SL, Eq.~\eqref{singlelens}) and 
large clusters limits (LC, Eq.~\eqref{evademicrov2}) 
and Lyman-$\alpha$ constraints (${\rm L}\alpha$, Eq.~\eqref{Lyboundv2}).
The hatched regions are ruled out by either microlensing (red) and/or Lyman-$\alpha$ (blue) limits, indicating that the combination of these two experiments alone entirely constrains the relevant parameter space.
Finally, for reference, we show with the vertical dashed line the PBH average distance if they were Poisson distributed at formation, while the dot-dashed diagonal line refers to condition~\eqref{accretion} for coherent accretion.
	}
	\label{figconstraint}
\end{figure}

As these clusters are still very compact, 
it is easy to check that they are not destroyed by tidal effects coming from interactions with the surrounding environment (see the SM for details). 
On the other hand, their smaller physical size makes it easier for PBH clusters to dynamically evaporate~\cite{1987gady.book.....B}. 
The minimal number of PBHs in the cluster to avoid evaporation within the age of the universe can be translated into a constraint on the PBH initial correlation function (see SM for details)
\be
\label{xievap}
\text{Eva}: \quad \xi_0 \lesssim 1.7 \cdot 10^{-3} f_\PBH \lp \frac{200}{C}\rp^{1/2}
\lp \frac{M_\odot}{M_\PBH} \rp
\lp \frac{r_\text{\tiny cl}}{\rm kpc} \rp^{3} .
\ee
This condition, shown as a red solid line in Fig.~\ref{figconstraint}, covers a large parameter space for the initial PBH clustering.
Consequently, initial strong clustering $\xi_0 \gg 1$ enhances cluster evaporation when compared to Poisson initial conditions.
More general considerations about PBH clustering modelling can be found in the SM.

\vskip 0.5cm
\noindent
\noindent{{\bf{\it Microlensing constraints.}}}
Microlensing surveys provide a powerful probe to constrain PBHs in a wide range of masses. 
They search for the temporary amplification of distant sources like stars due to the passage of a compact object near the line-of-sight~\cite{Paczynski:1985jf}. Several constraints were set on the abundance of PBHs in the Milky Way halo. Examples are given by observations of M31 using Subaru HSC~\cite{Niikura:2017zjd,Croon:2020ouk}, which set a limit on planetary and sub-planetary PBH masses, while EROS, MACHO and OGLE surveys of the Magellanic Clouds constrain stellar and planetary PBH masses~\cite{EROS-2:2006ryy, Macho:2000nvd, Blaineau:2022nhy, Wyrzykowski:2015ppa}.
Overall, stellar microlensing constrains impose $f_\PBH \lesssim 0.1$ in the mass range $10^{-10} M_\odot \lesssim M_\PBH \lesssim 10^3 M_\odot$, thus excluding PBHs for making up all of the DM in this mass range~\cite{Green:2020jor}.

These limits were derived assuming evenly distributed PBHs.
It was suggested that the inevitable clustering of PBHs induced by Poisson initial conditions could significantly relax these bounds~\cite{Clesse:2017bsw,Calcino:2018mwh,Carr:2019kxo}, as they would act as a single lens with a mass much larger than the one relevant for the microlensing surveys.
This result was disputed by Refs.~\cite{Petac:2022rio,Gorton:2022fyb}, which found that this criterion can only be satisfied either for very compact PBH clusters, that act as a point-like object, or for non-compact clusters containing a sufficiently large amount of objects, where individual PBHs can be resolved, neither of which are reached in initially Poisson scenarios.
On the other hand, initially clustered scenarios may avoid these bounds, provided that PBH clusters remain stable, \ie, satisfy Eq.~\eqref{xievap}.

The first condition is realised if the Einstein radius of the cluster is larger than its size, $R_\E (M_\cl) \gtrsim r_\text{\tiny b}$. For surveys of Magellanic clouds, the Einstein radius is~\cite{Petac:2022rio,Gorton:2022fyb}
\be
R_\E (M_\cl) \simeq 4.8 \cdot 10^{-5} {\rm pc} \, \sqrt{\frac{M_\text{\tiny cl}}{M_\odot}},
\ee
from which one extracts the condition for compact-enough PBH clusters to act as a single lens 
\be
\label{singlelens}
\text{SL}: \quad 
\xi_0 \gtrsim 18\, f_\PBH^{-1/9} \lp \frac{C}{200}\rp^{-2/9} \lp \frac{r_\text{\tiny cl}}{\rm kpc} \rp^{-1/3}.
\ee
This condition is shown as a magenta line in Fig.~\ref{figconstraint}, and it is meaningful only in the parameter space where cluster evaporation is not efficient enough (see the transition from the dashed to the solid line in Fig.~\ref{figconstraint}).

On the other hand, clusters which are not compact enough, but have a large amount of PBHs and act as a compilation of spatially correlated individual lenses, can still evade the microlensing limits. 
This occurs if the number of PBHs in clusters is larger than \cite{Gorton:2022fyb}
\be
\label{evademicro}
    N_\text{\tiny cl} \gtrsim 10^6 \lp \frac{M_\PBH}{M_\odot} \rp^{-1},
\ee
which requires
\be
\label{evademicrov2}
    \text{LC}: \quad \xi_0 \gtrsim 7.7 \cdot 10^3 f_\PBH^{-1} \lp \frac{r_\text{\tiny cl}}{\rm kpc} \rp^{-3}.
\ee
This bound is depicted in green in Fig.~\ref{figconstraint}.
Again, this holds only when PBH clusters do not evaporate efficiently.

\vskip 0.5cm
\noindent
\noindent{{\bf{\it Constraints from Lyman-$\alpha$ observations.}}}
We now briefly review bounds obtained from the Lyman-$\alpha$ forest 
and describe how they change if PBHs are initially clustered.

The Lyman-$\alpha$ forest is a series of absorption lines in the spectra of distant galaxies and quasars arising from the transitions of electrons from the ground state to the first excited state of the neutral hydrogen atom. As the light travels through multiple gas clouds with different redshifts, multiple absorption lines may be formed~\cite{Viel:2001hd, Viel:2005qj, Viel:2013fqw}.

Assuming that PBHs are initially Poisson distributed, Refs.~\cite{Afshordi:2003zb,Murgia:2019duy} have studied their impact on the Lyman-$\alpha$ observations. The PBH contribution to the matter linear power spectrum would be
$P_\PBH (k) = {f_\PBH^2}/{\overline n_\PBH}$, 
which enhances the standard $\Lambda$CDM spectrum producing a small-scale plateau. 
As the adiabatic contribution evolves as $k^{-3}$ at large $k$, the isocurvature term would be important only at scales relevant for Lyman-$\alpha$ observations. Ref.~\cite{Murgia:2019duy} found a $(2 \sigma)$ upper limit of the form
\be
    f_\PBH M_\PBH \lesssim 60\, M_\odot,
\ee
when a Gaussian prior on the reionization redshift is assumed. 
For large PBH abundances $f_\PBH = 1$, these constraints can be interpreted as limits on the PBH mass, while they become weaker for small abundances, up to values $f_\PBH \approx 0.05$ where seed effects could modify the predictions. Following Ref.~\cite{Murgia:2019duy}, we do not consider values smaller than $f_\PBH \lesssim 0.05$. 

Crucially, Lyman-$\alpha$ observations concern (at moderate redshifts $\sim 5$), comoving scales between ${\cal O}(10^{-1}\divisionsymbol 1)$ Mpc, which are much larger than the typical cluster scale $r_\cl$. 
At such large scales, clustered PBHs can be treated as compact objects with mass $M_\cl$ following a Poisson distribution in space with mean distance $R_\cl$.
This implies that, for strongly clustered PBHs, the Lyman-$\alpha$ bound translates into the condition
\be
\label{Lyboundv1}
    f_\PBH M_\text{\tiny cl} = f_\PBH N_\text{\tiny cl} M_\PBH \lesssim 60\, M_\odot \,.
\ee
It can be rewritten as
\be
\label{Lyboundv2}
    \text{L}\alpha: \quad \xi_0 \lesssim  0.5 f_\PBH^{-2} \lp \frac{r_\text{\tiny cl}}{\rm kpc} \rp^{-3}.
\ee
This is shown as a blue solid line in Fig.~\ref{figconstraint}. 
The bound is conservative as it does not for account the evolution of initially Poisson distributed PBH clusters under the action of self-gravity before redshift $z = 199$, at which Ref.~\cite{Murgia:2019duy} fixed their initial conditions. Including PBH clustering evolution~\cite{Inman:2019wvr,DeLuca:2020jug} would result in a larger power spectrum at scales of interest for the Lyman-$\alpha$ observations and therefore to a bound stronger than Eq.~\eqref{Lyboundv2}.

\vskip 0.5cm
\noindent
\noindent{{\bf{\it Other constraints in the stellar mass range.}}}
Strong initial PBH clustering affects constraints on $f_\PBH$ also from other independent observables. While a complete analysis of those is beyond the scope of the paper, we highlight the qualitative effect of clustering on bounds from CMB observations and GW measurements and quantify when these effects are expected to become relevant.

PBHs heavier than the stellar mass are bounded by accretion constraints from Planck data~\cite{Ali-Haimoud:2016mbv,Serpico:2020ehh}. 
Indeed, soon after the matter-radiation equality, heavy enough PBHs could start accreting baryonic particles from the surrounding medium. The resulting emission of ionizing radiation can alter the opacity of the gas in the period between recombination and reionization, thus affecting the CMB temperature and polarization fluctuations. 
In the conservative case, Planck data impose roughly $f_\PBH \lesssim (M_\PBH/10 M_\odot)^{-2}$ in the LVK mass range~\cite{Serpico:2020ehh}.

If PBHs are strongly clustered, these constraints can change. Compact PBH clusters could accrete as a coherent object if the accretion radii of individual PBHs overlap significantly, enhancing the accretion rate by at most a factor of $N_{\cl}$ when compared to the case where each PBH is an independent accretor~\cite{Lin:2007pc,Kaaz:2019wdi,Hutsi:2019hlw}. This occurs if the average PBH separation in the cluster, i.e. $r_\text{\tiny b}/N_\cl^{1/3}$, is smaller than the individual PBH Bondi radii $r_{B,\PBH} = G M_\PBH/v_\text{\tiny eff}^2$, in terms of their effective velocity, which we assume to be dominated by the virial velocity of PBHs in the cluster (see the SM),
and requires
\be
\label{accretion}
    \xi_0 \gtrsim 7.7 \cdot 10^{-3} f_\PBH^{-1} \lp \frac{M_\PBH}{M_\odot} \rp  \lp \frac{r_\text{\tiny cl}}{\rm kpc} \rp^{-3}.
\ee
Strong PBH clustering can tighten bounds coming from accretion at high redshifts in all scenarios with initial clusters characterised by $N_\text{\tiny cl}\gtrsim 3$, as illustrated in Fig.~\ref{figconstraint}. It is conceivable that coherent accretion into PBH clusters alone could close the LVK mass window for clustered PBH-DM, even though a detailed analysis is required.

Although initial PBH clustering enhances the formation of PBH binaries in the early universe~\cite{Raidal:2017mfl}, even relatively mild clustering can cause nearly all of these binaries to be disrupted~\cite{Raidal:2018bbj,Vaskonen:2019jpv,Jedamzik:2020ypm}. Nevertheless, disrupted binaries would still result in a merger rate too large to be consistent with LVK observations~\cite{LIGOScientific:2016aoc, LIGOScientific:2018mvr, LIGOScientific:2020ibl, LIGOScientific:2021djp}, excluding scenarios in which solar mass PBHs make up the entire DM~\cite{Vaskonen:2019jpv}. However, in extreme cases, all PBH binaries formed in the early universe may merge well before the present. 
In this case, a large number of very compact PBH binaries would be produced in the early universe, but would not contribute to the present merger rate observed by LVK due to their short coalescence times. 
Nonetheless, they could still give rise to a potentially observable stochastic gravitational wave background. Only binaries forming in the later PBH clusters, likely through 3-body interactions~\cite{Franciolini:2022ewd}, would contribute to the present BH-BH merger rate. These effects indicate a very different GW phenomenology of PBH binaries when compared to the Poisson case. 

To roughly estimate the minimum value of the initial PBH correlation for such a scenario, consider a circular PBH binary with its initial semi-major axis given by the mean separation in the cluster, $r_\text{\tiny b}/N_\text{\tiny cl}^{1/3}$, and demand that it coalesces within a Hubble time, that is 
\be
    \xi_0 \gtrsim 5.8 \cdot 10^3 \lp \frac{C}{200} \rp^{-1/4} \lp \frac{M_\PBH}{M_\odot} \rp^{-5/16}.
\ee
If this condition is met, all initial PBH binaries are expected to have merged before the present day.

\vskip 0.5cm
\noindent
\noindent{{\bf{\it Discussion and conclusions.}}}
All our simple considerations about initially clustered PBH scenarios induced by a large initial two-point correlation leads to Fig.~\ref{figconstraint}, showing that existing constraints cannot be avoided for $f_\PBH = 1$. The same conclusion is valid also for $f_\PBH = 0.1$, as we show in the SM. 
The main reason is simple: too light PBH clusters would evaporate and standard microlensing constraints hold; 
on the other hand, although sufficiently heavy PBH clusters may be stable and escape microlensing limits, they would inevitably induce large scale perturbations incompatible with the Lyman-$\alpha$ observations.

In more detail, the red region in Fig.~\ref{figconstraint} covers the parameter space where PBH clusters do not evade the microlensing bounds because of evaporation. 
In the complementary region, PBH clusters would be stable and may avoid microlensing bounds either by acting as a single lens or by being large enough. 
However, the blue region indicates that in such parameter space Lyman-$\alpha$ bounds would apply. 
Therefore, the combination of those constraints alone prevents stellar-mass primordial black holes
to be the entirety of the dark matter, even if one considered non-standard initial conditions with PBHs strongly clustered already at their formation.

We checked that the conclusions summarised in Fig.~\ref{figconstraint} do not change assuming a different PBH mass in the stellar mass range. While for smaller PBH masses the parameter space bounded by requiring a minimum number of PBHs in the cluster would shift to smaller $r_\cl$, it is harder for cluster evaporation to occur, thus extending the region where microlensing limits apply. 
On the other hand, for larger PBH masses, the consistency bound $N_\cl \gtrsim 3$ would shift to the right and evaporation more easily occurs. The bound from Lyman-$\alpha$, however, does not depend on the $M_\PBH$ explicitly, therefore leaving our results unaffected.

Our conclusions hold independently of the specific form of the two-point correlator on small scales $r < r_\cl$ (which determines the PBH profile inside the clusters) as long as the clusters are approximately Poisson distributed on scales much larger than $r_\cl$. 
For the Lyman-$\alpha$ and the single lens condition, a sufficiently compact PBH cluster behaves as a massive compact halo object (MACHO) independently of its composition. Importantly, this implies that our results are independent of the exact distribution of PBH masses making up this object.
Finally, although our conclusions are derived assuming a monochromatic spectrum for PBH cluster masses, the case of a wider cluster mass spectrum can be addressed using existing methods for wide PBH mass distributions~\cite{Carr:2017jsz, Bellomo:2017zsr} but with PBHs now replaced by compact PBH clusters. In particular, the relevant Lyman-$\alpha$ constraints for extended mass functions were studied in Ref.~\cite{Murgia:2019duy}.
In all, we expect our conclusions to withstand reasonable variations of the initial PBH two-point function, the distribution of the masses of PBHs and their initial clusters.

While our considerations rule out models of PBH formation with initial clustering (see for instance Refs.~\cite{Belotsky:2018wph,Ding:2019tjk,Atal:2020igj}), we stress that our findings close the window for stellar-mass PBHs as dark matter. 

\vskip 0.5cm
\noindent
\noindent{{\bf{\it Acknowledgments.}}}
We thank M. Biagetti and M. Viel for interesting discussions and A. Green and K. Jedamzik for useful feedbacks on the draft.
V.DL. and A.R. are supported by the Swiss National Science Foundation (SNSF), project {\sl The Non-Gaussian Universe and Cosmological Symmetries}, project number: 200020-178787.
G.F. acknowledges financial support provided under the European
Union's H2020 ERC, Starting Grant agreement no.~DarkGRA--757480 and under the MIUR PRIN programme, and support from the Amaldi Research Center funded by the MIUR program ``Dipartimento di Eccellenza" (CUP:~B81I18001170001). This work was supported by the EU Horizon 2020 Research and Innovation Programme under the Marie Sklodowska-Curie Grant Agreement No. 101007855.
H.V. is supported by the Estonian Research Council grants PRG803 and MOBTT5, and the EU through the European Regional Development Fund CoE program TK133 “The Dark Side of the Universe”.

\begin{center}
    
{{\bf{\bf Supplemental Material}}}

\end{center}

\vskip 0.5cm
\noindent
\noindent{{\bf{\it A simple model of initially clustered PBHs.}}}
To improve our intuition about initially clustered PBHs, it is instructive to build an explicit model of such a scenario and use it to derive the statistical characteristics of the spatial distribution resulting from this construction. We will model an initial spatial distribution from the following assumptions: 
{\it i)} PBHs are initially distributed in clusters described by a profile
    \be\label{eq:P_C}
        P_\cl(\vec{x}|\vec{x}_\cl) = n_\cl(|\vec{x} - \vec{x}_\cl|)/N_\cl,
    \ee
where $\vec{x}_\cl$ denotes the central position of the cluster and $P_\cl(\vec{x}|\vec{x}_\cl)$ is normalized to one to allow for a probabilistic interpretation. For simplicity, each cluster is assumed to contain $N_\cl$ PBHs and we assume a monochromatic mass function for PBHs. $n_\cl$ can be interpreted as the number density profile. We stress, however, that $n_\cl$ is the initial profile \emph{before} the gravitational collapse of the cluster and it can considerably differ from the profile of the gravitationally bound system.
{\it ii)} The PBH positions inside the clusters are distributed independently according to Eq.~\eqref{eq:P_C}.
{\it iii)} The clusters follow a uniform Poisson distribution with density 
$\overline n_\text{\tiny cl} = \overline n_\PBH/N_\cl$.
The last two assumptions simply mean that neither the substructure of the clusters nor the spatial distribution of the clusters generates non-Poisson contributions, so the initial (non-Poisson) clustering is completely captured by the density profile \eqref{eq:P_C}.

First, within a single cluster, the probability to find a PBH at position $\vec x$ if there is another PBH at position $\vec y$ is given by the convolution of cluster profiles
\begin{eqnarray}
    P_1(\vec{x}|\vec{y}) 
    &=& \frac{P_1(\vec{x},\vec{y})}{P_1(\vec{y})}
    = \frac{\int \d^3 x_\cl \, P_{\cl}(\vec{x}|\vec{x}_{\cl}) P_{\cl}(\vec{y}|\vec{x}_{\cl}) p(\vec{x}_{\cl})}{\int \d^3 x_\cl \, P_{\cl}(\vec{y}|\vec{x}_{\cl}) p(\vec{x}_{\cl})}\nonumber\\
    &=& N_\cl^{-2}\int \d^3 x_\cl \, n_\cl(|\vec{x}-\vec{y}-\vec{x}_{\cl}|) n_\cl(|\vec{x}_{\cl}|).
\end{eqnarray}
Note that $P_1(\vec{x})$ is constant if the distribution of clusters $p(\vec{x}_{\cl})$ is uniform.
To obtain the conditional probability $P(\vec{x}|\vec{y})$ for the clustered distribution we must include the possibility that the PBH at $\vec{x}$ is in a different cluster than the PBH at $\vec{y}$. To this aim, consider a large but finite volume $U$ containing $N_U$ clusters so that the probability of finding both PBHs in the same cluster is $(N_\cl -1)/(N_U N_\cl-1)$. Therefore,
\bea
    P(\vec{x},\vec{y}) 
&    = \frac{N_\cl -1}{N_U N_\cl-1} P_1(\vec{x},\vec{y}) \\
&    + \frac{(N_U -1)N_\cl}{N_U N_\cl-1} P_1(\vec{x})P_1(\vec{y}),
\eea
where $P_1(\vec{x}) = 1/U$. Taking the limit $U \to \infty$ with $N_U/U = \bar n_\cl$ fixed and multiplying with the total number of PBHs, $N_{\cl} N_U$, gives the density of PBHs surrounding a PBH located at $y$,
\bea
    n_{\PBH}(\vec{x}|\vec{y}) 
&    = \bar n_{\PBH} + (N_\cl-1) P_1(\vec{x}|\vec{y}) \\
&    \equiv \bar n_{\PBH} \left[1 + \xi_{\PBH}(|\vec{x}-\vec{y}|) \right],
\eea
so the non-Poisson component arises purely from the halo profile as stated above. As expected, the distribution reduces to Poisson if $N_\cl = 1$. Importantly, 
\be
    \overline{n}_\PBH \int \d^3 x  \, \xi_\PBH = N_\cl - 1,
\ee
since $P_1(\vec{x}|\vec{y})$ is a probability density and thus normalized to unity. In the limit $N_\cl \gg 1$, we obtain Eq.~\eqref{eq:N_cl}. An analogous derivation gives also Eq.~\eqref{eq:delta_2P} for the two-point function of the density contrast.

For instance, consider initial clusters with a Gaussian profile
\be
    n_\cl(r) 
    = n_{\star} e^{-r^2/(2r_{\cl}^2)}
    , \quad
    n_{\star} \equiv N_\cl/(\sqrt{2\pi}r_{\cl})^3.
\ee
Then, the two-point function reads
\be
    \xi_{\PBH}(r) = \frac{\delta_{\star}}{2^{3/2}} e^{-r^2/(4r_{\cl}^2)},
\ee
where $\delta_{\star} \equiv n_{\star}/\overline n_{\PBH}$ denotes the central density enhancement in the clusters, being $n_{\star}$ the number density in the center of clusters. For $r\lsim r_\cl$, one recovers Eq.~\eqref{corr_template} where $\xi_0\simeq \delta_\star/2^{3/2}$.

A few conclusions can be drawn from this simple picture: i) Non-trivial two-point functions can be realized/constructed by uniformly distributing halos with identical profiles. However, the relation is not one-to-one, so not all two-point functions correspond to specific cluster profiles. ii)  $\xi_{\PBH}$ is wider and less peaked than the initial cluster profile. iii)  In this construction, the clusters collapse after re-entry and, as they are statistically identical, the resulting cluster mass function is essentially monochromatic peaking at $M_{\cl} = N_{\cl} M_{\PBH}$ until the clusters begin to merge in the late universe following the evolution seeded by their initial Poisson distribution~\cite{Inman:2019wvr,DeLuca:2020jug}. We stress that the constructed spatial distribution is not Gaussian and thus the two-point function only provides a partial description of it. Especially, as it is possible to construct explicit initial distributions by assuming a wide distribution for $N_\cl$ while keeping $\xi_{\PBH}$ fixed, it follows that $\xi_{\PBH}$ does not determine the cluster mass function uniquely and scenarios with different cluster mass functions must be distinguished by higher order correlators.

\begin{figure*}[t!]
	\centering
\includegraphics[width=0.49\textwidth]{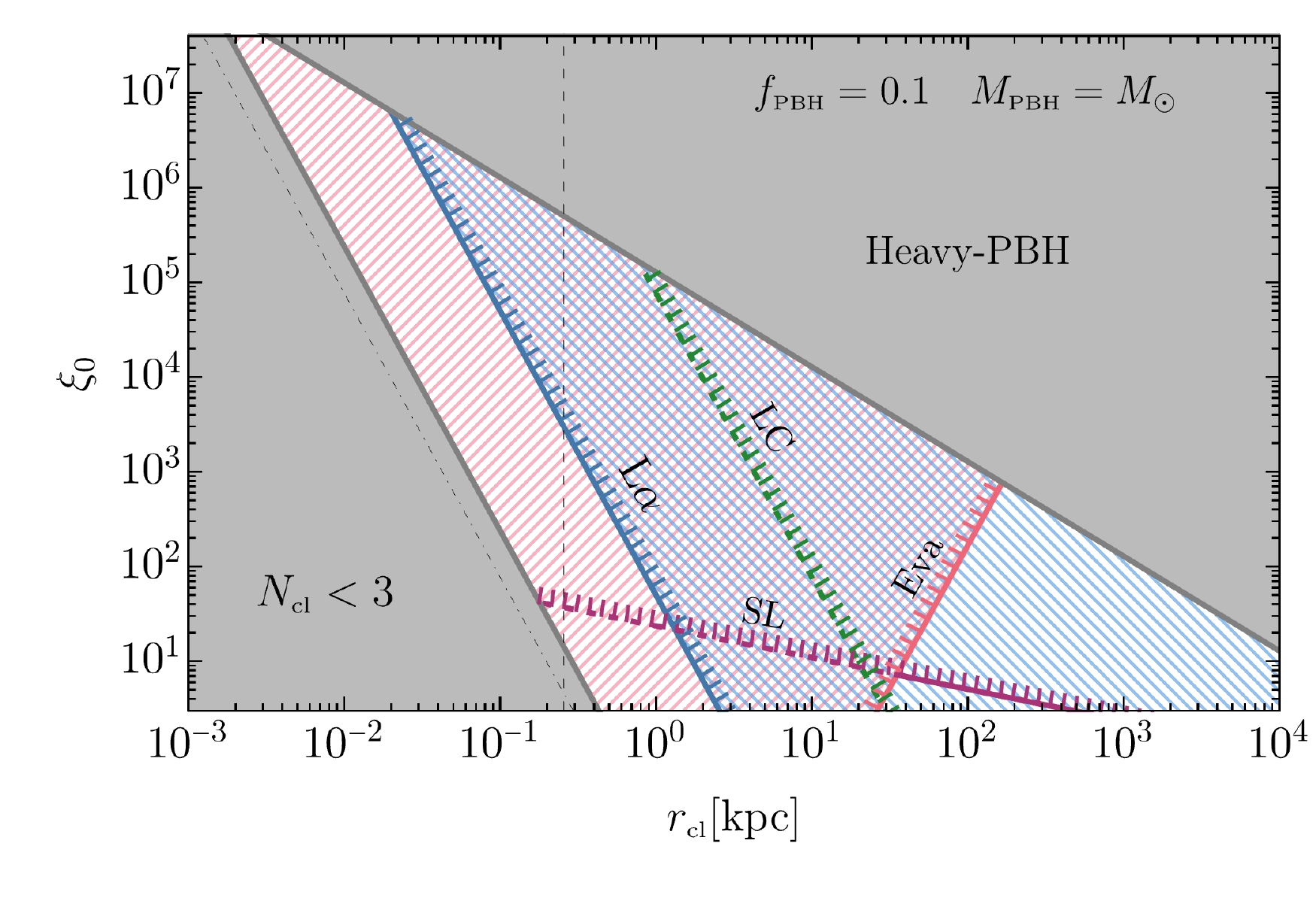}
\includegraphics[width=0.49\textwidth]{Plots/Constraint_plane_fPBH_1.pdf}
	\caption{
	Same of Fig.~\ref{figconstraint} comparing $f_\PBH = 0.1$ (left panel) and $f_\PBH = 1$ (right panel). 
	}
	\label{figconstraint2}
\end{figure*}

\vskip 0.5cm
\noindent
\noindent{{\bf{\it Gravitational collapse of initial PBH clusters.}}}
Consider the evolution of an expanding clump of PBHs into a gravitationally bound PBH cluster during radiation domination. We can assume that $N_{\cl} \gg 2$ as otherwise the collapse would produce a short-lived N-body system. 
Assuming that the average density contrast within the initial comoving cluster radius $r_{\cl}$ is approximately $\xi_0 \gg 1$, the scale factor at decoupling is approximately
\be
    a_{\cl} = a_{\eq}/\xi_0,
\ee
where $a_{\eq}$ indicates the scale factor at the epoch of matter-radiation equality.
In principle, the interior region that has the highest density contrast can decouple before the outer shells. This can be used to convert the initial density profile \eqref{eq:P_C} into a profile of the bound system. However, as the density profile is expected to evolve further during relaxation, reliable estimates for it will likely require a numerical treatment. The average density of such clusters is roughly \cite{Kolb:1994fi}
\be
    \rho_{\cl} 
    = C \rho_{\rad} (a_{\cl})
    = C \rho_{\eq} \xi_0^4,
\ee
where $C = {\cal{O}}(1 \divisionsymbol 10^2)$ is a constant of proportionality, which may depend on the initial cluster profile \eqref{eq:P_C}. The proper final radius of the bound cluster is
\begin{align}
\label{finalradius}
    r_\text{\tiny b} 
    & = \lp {\frac{3 M_{\cl}}{4\pi \rho_{\cl}}} \rp ^{1/3}
    \simeq  0.05\, {\rm pc} \, \xi_0^{-\frac{4}{3}} \left(\frac{M_{\cl}}{C M_\odot}\right)^{\frac{1}{3}} 
    \nonumber \\
    & \simeq 4 \cdot 10^{-2} {\rm pc} \,
    f_\PBH^{1/3} \xi_0^{-1}
    \lp \frac{C}{200} \rp^{-1/3}
    \lp \frac{r_\text{\tiny cl}}{\rm kpc} \rp,
\end{align}
while the virial PBH velocity in the cluster is
\begin{align}
\label{virial}
v_\text{\tiny b} &=
\lp \frac{G N_\text{\tiny cl} M_\PBH }{r_\text{\tiny b}} \rp^{1/2} \nonumber \\
& \simeq 
3.6\, {\rm km/s}\, 
f_\PBH^{1/3} \xi_0 \lp \frac{C}{200}\rp^{1/6}   \lp \frac{r_\text{\tiny cl}}{\rm kpc} \rp.
\end{align}
If we compare $r_\text{\tiny b}$ to the one found with Poisson initial conditions, \ie, $r_\text{\tiny b} \simeq 10^{-2} N_\cl^{5/6} (M_\PBH/M_\odot)^{1/3} {\rm pc} $~\cite{Gorton:2022fyb}, 
we see that initial clustering produces consistently smaller (and more compact) PBH clusters. Importantly, this modifies the evaporation timescales as well as lensing properties of the cluster, as discussed in the main text.

\vskip 0.5cm
\noindent
\noindent{{\bf{\it Evaporation of PBH clusters.}}}
The evaporation time of a system of $N_\text{\tiny cl}=M_\text{\tiny cl}/M_\PBH$ PBHs clustered in a region of size $r_\text{\tiny b}$ and subject to the gravitational force is given by~\cite{1987gady.book.....B}
\begin{align}
t_{\text{\tiny{ev}}}
& = 14 \frac{N_\text{\tiny cl}}{\log N_\text{\tiny cl}} \frac{r_\text{\tiny b}}{v_\text{\tiny b}} \nonumber \\
& \simeq \frac{10^{11} {\rm yr}}{\log N_\text{\tiny cl}} \left( \frac{N_\text{\tiny cl}}{10^6} \right)^{1/2} \left(  \frac{M_\PBH}{M_\odot}\right)^{-1/2}  \left( \frac{r_{\text{\tiny b}}}{{\rm pc}} \right)^{3/2}.
\end{align}
Substituting Eq.~\eqref{finalradius} in the evaporation timescale, one can extract the minimum number of PBHs in the cluster necessary to avoid evaporation in the age of the universe as
\be
N_\text{\tiny cl} \gtrsim 7.9 \cdot 10^4 
\xi^2_0
\left (\frac{C}{200}\right)^{1/2} .
\ee
This result shows that, when PBHs are initially clustered, a much larger number of PBHs in the clusters are necessary to avoid evaporation from these systems.
This condition can be translated into a constraint in the clustered PBH parameter space. Using   $N_\cl \sim \overline{n}_\PBH r^3_\cl \xi_0$, we finally get Eq.~\eqref{xievap}.
We infer therefore that large initial PBH clustering $\xi_0 \gg 1$ gives rise to compact clusters which evaporate more easily than clusters born from Poisson initial conditions, making microlensing constraints more difficult to be evaded.

\vskip 0.5cm
\noindent
\noindent{{\bf{\it Tidal disruption of PBH clusters.}}}
Encounters between two PBH clusters may induce tidal disruption of the clusters. 
Under the ``distant-tide'' approximation, applicable when the separation between these objects is significantly larger than their physical size, an encounter increases the cluster's internal energy, and may eventually disrupt it.
The characteristic timescale of this process is given by~\cite{1958ApJ...127...17S}
\begin{align}
t^\text{\tiny b-b}_{\text{\tiny{tidal}}} &= \frac{1}{8\pi G M_\text{\tiny cl} \overline{n}_\text{\tiny cl}} \frac{v^\text{\tiny b-b}_\text{\tiny rel}}{r_\text{\tiny b}} \nonumber \\
&\simeq 1.6 \cdot 10^{7} {\rm Gyr} \, f_\PBH^{-4/3} \xi_0 \lp \frac{C}{200}\rp^{1/3}  \lp \frac{v^\text{\tiny cl-cl}_\text{\tiny rel}}{{\rm km/s}} \rp  \left( \frac{r_{\text{\tiny cl}}}{{\rm kpc}} \right)^{-1},
\end{align}
in terms of the relative velocity $v^\text{\tiny b-b}_\text{\tiny rel}$ between the two PBH clusters. 
The timescale for disruption becomes significantly larger when clusters are more compact, \ie, for larger $\xi_0$ and $C$ and/or smaller $r_\text{\tiny cl}$.
As one can appreciate, using their typical values relevant in our discussion (see, \eg, Fig.~\ref{figconstraint}), this process occurs on timescales much larger than the current age of the universe.

\vskip 0.5cm
\noindent
\noindent{{\bf{\it Summary of main bounds.}}} The main bounds found in the letter are summarized in Table.~\ref{Tablebound}. A comparison of our results for $f_\PBH = 0.1$ (left panel) and $f_\PBH = 1$ (right panel) is shown in Fig.~\ref{figconstraint2}.

\begin{table}[!h!]
{\renewcommand{\arraystretch}{1.6}
\label{Tablebound}
\begin{tabular}{|c|c|}
\hline
\hline
\multicolumn{2}{|c|}{
Consistency of the clustered scenario}
\\ 
\hline
\hline
$N_\cl \gtrsim 3$                  
& $\xi_0 \gtrsim 2.3 \cdot 10^{-2} f_\PBH^{-1}  \lp \frac{M_\PBH}{M_\odot} \rp
\lp \frac{r_\text{\tiny cl}}{\rm kpc} \rp^{-3}$
\\ \hline
Heavy-PBH                  
& $\xi_0 \lesssim 6 \cdot 10^4\, f_\PBH^{-1/3} \lp \frac{C}{200}\rp^{-1/6} \lp \frac{r_\cl}{\rm kpc} \rp^{-1}$
\\ \hline
\hline
\multicolumn{2}{|c|}{
Conditions to evade individual bounds}
\\ \hline
\hline
Evaporation              
& $\xi_0 \lesssim 1.7 \cdot 10^{-3} f_\PBH \lp \frac{200}{C}\rp^{1/2}
\lp \frac{M_\odot}{M_\PBH} \rp
\lp \frac{r_\text{\tiny cl}}{\rm kpc} \rp^{3}$
\\ \hline
Single Lens                  
& $\xi_0 \gtrsim 18\, f_\PBH^{-1/9} \lp \frac{C}{200}\rp^{-2/9} \lp \frac{r_\text{\tiny cl}}{\rm kpc} \rp^{-1/3}$
\\ \hline
Large Clusters                  
& $\xi_0 \gtrsim 7.7 \cdot 10^3 f_\PBH^{-1} \lp \frac{r_\text{\tiny cl}}{\rm kpc} \rp^{-3}$
\\ \hline
Lyman-$\alpha$                   
& $\xi_0 \lesssim  0.5 f_\PBH^{-2} \lp \frac{r_\text{\tiny cl}}{\rm kpc} \rp^{-3}$
\\ 
\hline
\hline
\end{tabular}
\caption{Summary the main bounds on the correlation function that enter in Fig.~\ref{figconstraint} and Fig.~\ref{figconstraint2}.}}
\end{table}

\bibliography{draft}

\end{document}